\documentclass{aa}  

\usepackage{ulem}
\renewcommand{\arraystretch}{1.1}
\usepackage{color}
\usepackage{array}
\usepackage{url}
\usepackage[bottom]{footmisc}
\usepackage{xcolor}
\usepackage{cuted}   
\usepackage{capt-of} 
\usepackage{needspace}

\usepackage{graphicx}
\usepackage{multirow}
\usepackage{booktabs}
\usepackage{caption}
\usepackage{subcaption}
\usepackage{booktabs}
\usepackage{adjustbox}
\usepackage[toc,page]{appendix}
\usepackage{amsmath}
\usepackage{nccmath}
\usepackage{arydshln} 
\bibpunct{(}{)}{;}{a}{}{,}
\include{definitions}
\newcommand{\Ngc}{{NGC 5055 ULX X-1}\xspace}
\newcommand{\ngc}{{NGC~5055}\xspace}

\usepackage{siunitx}

\newcommand{\chan}{\textit{Chandra}\xspace}
\newcommand{\swift}{\textit{Swift}\xspace}
\newcommand{\xmm}{XMM--\textit{Newton}\xspace}
\newcommand{\nustar}{\textit{NuSTAR}\xspace}

\newcommand{\rosat}{\textit{ROSAT}\xspace}

\usepackage{txfonts}
\usepackage[breaklinks=true]{hyperref}
\hypersetup{unicode=true, colorlinks=true, linkcolor=[rgb]{0.53, 0.15, 0.34}, citecolor=[rgb]{0.06, 0.2, 0.65}, filecolor=[rgb]{1.0, 0.13, 0.32}, urlcolor=[rgb]{0.53, 0.15, 0.34}}

\begin{document} 

\title{The hard ultraluminous state of NGC 5055 ULX X-1}

   \author{N. Cruz-Sanchez\inst{1}, 
   E. A. Saavedra\inst{2,3},   
   F. A. Fogantini\inst{4},
   F. Garc\'{\i}a\inst{1,4} \and
   J. A. Combi\inst{1,4} 
   }

    \institute{
    Facultad de Ciencias Astron\'omicas y Geof\'isicas, Universidad Nacional de La Plata, B1900FWA La Plata, Argentina \and 
    Instituto de Astrof\'isica de Canarias (IAC), V\'ia Láctea, La Laguna, E-38205, Santa Cruz de Tenerife, Spain \and
    Departamento de Astrof\'isica, Universidad de La Laguna, E-38206, Santa Cruz de Tenerife, Spain \and
    Instituto Argentino de Radioastronom\'ia (CCT La Plata, CONICET; CICPBA; UNLP), C.C.5, (1894) Villa Elisa, Argentina }

 
   \abstract
   {
We present the results of the first broadband X-ray analysis of the ultraluminous X-ray source \Ngc, combining simultaneous data from \xmm and \nustar missions, with a combined exposure time of $\sim$100\,ks across the 0.3--20\,keV energy range. The source exhibits a stable flux across the entire exposure with no detectable pulsations by any instrument on their X-ray light curves, placing pulsed-fraction upper limits of 10\% and 32\% for \xmm and \nustar, respectively. The X-ray spectrum is dominated by two thermal components consistent with the emission from an accretion disk, and shows a weak high-energy tail above 10~keV, with no statistical requirement for an  additional nonthermal component.  The unabsorbed $0.3-20$~keV luminosity is ${\sim}2\times10^{40}$~erg\,s$^{-1}$, evidencing the ULX nature of the source. The parameters obtained from spectral modeling are consistent with the hard ultraluminous state. Despite the fact that a neutron-star accretor cannot be ruled out by the available data, under the assumption that the compact object in \Ngc is a black hole accreting through a geometrically thick, radiation-pressure-supported disk that drives an optically thick wind, we constrained its putative mass to $11-26\,M_\odot$.
}

   \titlerunning{The hard ultraluminous state of NGC 5055 ULX X-1}
   \authorrunning{Cruz-Sanchez et al.}
   \maketitle
%
 \section{Introduction}
Ultraluminous X-ray sources (ULXs) are extragalactic X-ray sources distinguished by their isotropic luminosities exceeding 10$^{39}$~erg~s$^{-1}$, surpassing the Eddington limit for stellar-mass black holes \citep[BHs;][]{1989ARA&A..27...87F, 2000ApJ...535..632M, 2011NewAR..55..166F}. These sources provide a unique laboratory for studying accretion processes under extreme conditions, whether through supercritical accretion onto stellar-mass compact objects or sub-Eddington accretion onto intermediate-mass black holes (IMBHs) \citep{1999ApJ...519...89C, 2000ApJ...535..632M, 2003ApJ...585L..37M, Strohmayer_2009_APJ, Fabrika_2021}. Although early interpretations invoked IMBHs to explain ULX luminosities, the detection of X-ray pulsations has confirmed that many ULXs harbor neutron stars (NSs). Additional systems are consistent with super-Eddington accretion onto stellar-mass BHs, favoring stellar-mass compact objects over IMBHs \citep{Fabrika_Mescheryakov_2001, King_2001, Mushtukov_2017, Fabrika_2021}.

Simultaneous observations with X-ray telescopes such as \xmm and \nustar~have enabled a more precise characterization of the spectral and temporal properties of ULXs. These studies have revealed a diversity of spectral states, commonly classified -- in the soft X-ray band only (0.3--10~keV), as originally defined by \citet{Sutton2013MNRAS} -- into three categories: the soft ultraluminous (SUL) state, the hard ultraluminous (HUL) state, and the broadened disk (BD) state \citep{Sutton2013MNRAS,Pintore_2020,Walton_2020}. In the SUL state, the spectrum is dominated by a cool thermal component ($kT<0.5$~keV) and a soft power-law index ($\Gamma>2$), while the HUL state is characterized by a hard power-law index ($\Gamma<2$) regardless of the thermal component. By contrast, the BD state features a hotter thermal component ($kT>0.5$~keV) dominating the spectrum, with a flux ratio between the power-law and disk components (F$_{\mathrm{PL}}$/F$_{\mathrm{disk}}$) $<5$ in the 0.3--1~keV range \citep{Sutton2013MNRAS, Walton_2020}. These classifications reflect the complex interactions between the accretion disk, the corona, and the outflows generated under supercritical accretion conditions.

A key aspect in the study of ULXs is the presence of NS, both pulsating and non-pulsating. The detection of coherent pulsations in several ULXs, such as M82 X-2 \citep{Bachetti_2014}, has confirmed the presence of highly magnetized NSs in these systems, now classified as pulsating ULXs (PULXs). These systems challenge the theoretical limits of super-Eddington accretion, as NSs can achieve luminosities hundreds of times their Eddington limit without being disrupted \citep{Mushtukov_2017,Fabrika_2021}. Furthermore, it has been observed that the spectra of PULXs are virtually indistinguishable from those of other ULXs, suggesting that a significant fraction of ULXs may be powered by NSs, even in the absence of detectable pulsations \citep{King_2023}. 

The spectral properties of ULXs differ markedly from those of Galactic black hole binaries (BHXBs). While BHXBs typically exhibit spectral curvature at energies above 10~keV, ULXs show a soft excess and spectral curvature below 10~keV, indicating the presence of a modified accretion disk and an optically thick corona \citep{Gladstone_2009, Sutton2013MNRAS}. While this was the prevailing interpretation at the time, more recent models favor an inner super-Eddington accretion disk surrounded by an outer disk/photosphere and powerful, radiatively driven winds \citep[e.g.,][]{Middleton2015MNRAS.454.3134M, Pinto2017MNRAS.468.2865P, Kosec2018MNRAS.479.3978K, Walton2018MNRAS.473.4360W}. Within this framework, the X-ray spectra are naturally described by multiple thermal components (sampled from the inner disk and the wind-photosphere) together with a high-energy excess above $\sim$10~keV, which is the approach adopted in this work \citep[e.g.,][]{Walton2018MNRAS.473.4360W}. Models of disk inflation and optically thick winds predict the formation of outflows, whose presence has been confirmed through the detection of blueshifted emission lines in the spectra of several ULXs \citep{2016Natur.533...64P, 2023A&A...671A...9A}. These outflows not only modulate the observed spectrum but also play a crucial role in regulating the accretion flow and the stability of the system.

\Ngc is a ULX candidate located in the spiral galaxy M63, at a distance of 9.04 Mpc \citep{Tully_2013,McQuinn_2017, Karachentsev_2020}. The source was first detected with \rosat \citep{10.1046/j.1365-8711.2000.03384.x} and has since been observed multiple times with \chan, \xmm, and \swift, consistently showing high X-ray luminosities up to $2.3 \times 10^{40}$ erg s$^{-1}$ \citep{Swartz_2011}. A systematic analysis of these archival observations was conducted by \citet{2020A&A...642A..94M}, who modeled the spectra using combinations of a multicolor disk (MCD) with either a power law or a thermal Comptonization component, while also testing a slim disk model. Their results revealed a spectral transition between soft and hard states, minimal short-term variability, a possible inverse correlation between luminosity and inner disk temperature, and indications of super-Eddington accretion accompanied by mild geometric beaming. However, the limited exposure times ($<30$~ks) and non-simultaneity of the observations limited the ability to constrain the spectral shape and its evolution.

Because of their high luminosities and complex spectral behavior, ULXs require broadband, temporally coordinated observations to constrain their emission mechanisms properly. Simultaneous data from \xmm and \nustar~provide complementary spectral coverage -- \xmm being sensitive to the soft X-ray regime and \nustar~to the hard X-rays -- allowing for robust modeling of both the thermal and high-energy components. This is particularly important to distinguish between BH and NS accretors, whose spectral signatures can otherwise appear degenerate in narrow energy bands.

Despite its persistent brightness, \Ngc (hereafter \ngc) has received limited attention. Previous studies were based on isolated observations with either \chan or \xmm at different epochs, each having exposure times shorter than $30$~ks \citep{2020A&A...642A..94M}, limiting spectral coverage and temporal coherence. In this paper, we present the first simultaneous broadband analysis of \ngc using \xmm and \nustar, with a combined exposure exceeding $100$~ks. This coordinated dataset enables a detailed investigation of both the continuum shape and potential variability, offering new insights into the nature of the accreting object.

This paper is structured as follows. In \hyperref[sec:datareduction]{Sect.~\ref{sec:datareduction}}, we describe the observational data and the reduction procedures applied to both \xmm and \nustar~observations. In \hyperref[sec:results]{Sect.~\ref{sec:results}} we present the main results of our timing and spectral analyses, including the search for pulsations and the broadband spectral characterization of \ngc. In \hyperref[sec:discussion]{Sect.~\ref{sec:discussion}}, we discuss the physical implications of these findings in the context of ULX accretion scenarios. Finally, in \hyperref[sec:conclusions]{Sect.~\ref{sec:conclusions}} we summarize our conclusions,   highlighting their relevance to the broader understanding of ULXs.

\section{Observations and data reduction} \label{sec:datareduction}
We analyzed simultaneous broadband X-ray data from the source \ngc, located at $\alpha$ = $13^h$ $15^m$ $19.69^s$ and $\delta = 42^{\circ}\,03^{\prime}\,02.1^{\prime\prime}$, obtained with \xmm (Newton Multi-Mirror X-ray Mission) and \nustar~\citep[Nuclear Spectroscopic Telescope Array;][]{2013ApJ...770..103H} observations. The calibration and data reduction processes for each observation are described in the following subsections.

\begin{table}[]
    \begin{center}
    \caption{\xmm and \nustar~ observations of \Ngc.}
    \label{tab:table_1}
    \begin{adjustbox}{max width=\columnwidth}
    \begin{tabular}{l c c c}
    \hline
    \hline
    Mission & Date & ObsID &   \\
    Camera & Exposure time (ks) & Good time$^{(a)}$ (ks)& Total counts\\
    \hline
    \xmm & 2021-12-06 & 0885180101 &  \\
    PN & 105.2 & 55.8 & 18438 \\
    M1 & 128.4 & 99.2 & 6394 \\
    M2 & 128.9 & 101.6 & 7489 \\
    \hline
    \nustar~ & 2021-12-05 & 30701003002 &  \\
    FPMA & 163.8 & 162.5 & 2470 \\
    FMPB & 162.3 & 161.7 & 2527 \\
    \hline
    \hline
    \end{tabular}
    \end{adjustbox}
    \end{center}
    \footnotesize{\textbf{Notes.} (a) The good time interval (GTI) for \xmm data is defined as the remaining exposure time obtained after removing  periods of high-energy background activity from the raw event files, while for \nustar~ data, the GTIs correspond to the remaining exposure time after processing the raw event data with the {\tt nupipeline} task.}
\end{table}

\subsection{\xmm} \label{xmm}

The \xmm observatory observed \ngc on December 6, 2021 (ObsID 0885180101) with an exposure time of 133~ks. The European Photon Imaging Camera (EPIC), which includes the pn \citep{Struder2001A&A...365L..18S} and MOS \citep{Turner2001A&A...365L..27T} detectors, operated in full-frame mode during the observation. 
Data reduction was performed using the \xmm Science Analysis System (SAS) software, version 21.0.0. Calibrated event lists were generated using the {\tt epproc} and {\tt emproc} tasks for the pn and MOS cameras, respectively. Periods of high-energy background flaring activity were identified and excluded using thresholds of 0.25 and 0.3 counts per second in the 10--12~keV energy band for the MOS cameras (MOS1 and MOS2, respectively) and 0.4 counts per second for the pn camera (for details and results see \hyperref[tab:table_1]{Table~\ref{tab:table_1}}). 
Source events were extracted from a circular region with a radius of 35 arcsec centered on \ngc, while background events were extracted from a nearby source-free region with a radius of 35 arcsec. Events were filtered using standard selection criteria: {\tt FLAG==0}, {\tt PATTERN{$\leq$}12} for MOS, and {\tt PATTERN{$\leq$}4} for pn. Light curves were extracted and corrected for background and exposure using the SAS task {\tt epiclccorr}, with barycentric corrections applied using the SAS task {\tt barycen}. The response matrices and auxiliary files were then generated using the SAS tasks {\tt rmfgen} and {\tt arfgen}, respectively. The final spectra were grouped to ensure a minimum of 20 counts per bin to facilitate spectral analysis using $\chi^2$ statistics.

\subsection{\nustar} \label{nustar}

\nustar~observed \ngc on December 05, 2021 (ObsID $30701003002$), with an exposure time of $\sim163$ ks. \nustar’s two co-aligned focal-plane modules, FPMA and FPMB, provide coverage in the 3--79 keV energy range. Data were reduced using \nustar~Data Analysis Software (NuSTARDAS) version 2.1.4, part of the HEASoft package (v.6.34), with calibration files from CALDB v.20241126. Calibrated event files were generated using the {\tt nupipeline} task, applying the South Atlantic Anomaly (SAA) filter with the parameters {\tt saacalc=3}, {\tt saamode=optimized} and {\tt tentacle=no}. This filtering resulted in a loss of $\lesssim0.9\%$ of the total exposure time (see \hyperref[tab:table_1]{Table~\ref{tab:table_1}}). The source events were extracted from a circular region with a radius of 55 arcsec centered on the position of \ngc, enclosing $\sim80\%$ of the point spread function (PSF). Background events were selected from a source-free region twice the size of the source, located on the same detector chip for FPMA, and in the opposite detector chip for FPMB, to ensure minimal contamination from the source and detector edges. Light curves and spectra were extracted using the {\tt nuproducts} task. Barycentric corrections were applied to the event files using the {\tt barycorr} task, with the {\tt nuCclock20100101} clock correction file and the {\tt JPL-DE200} Solar System ephemeris. Background-corrected light curves were produced for both FPMA and FPMB modules and then combined using the {\tt lcmath} task. The spectra were grouped to ensure a minimum of 20 counts per bin in the 3--78 keV energy range, allowing the use of $\chi^2$ statistics for spectral analysis.

\section{Analysis and results} \label{sec:results}

\subsection{Timing analysis} \label{lcur}

\begin{figure}
\centering
    \includegraphics[width=0.5\textwidth]{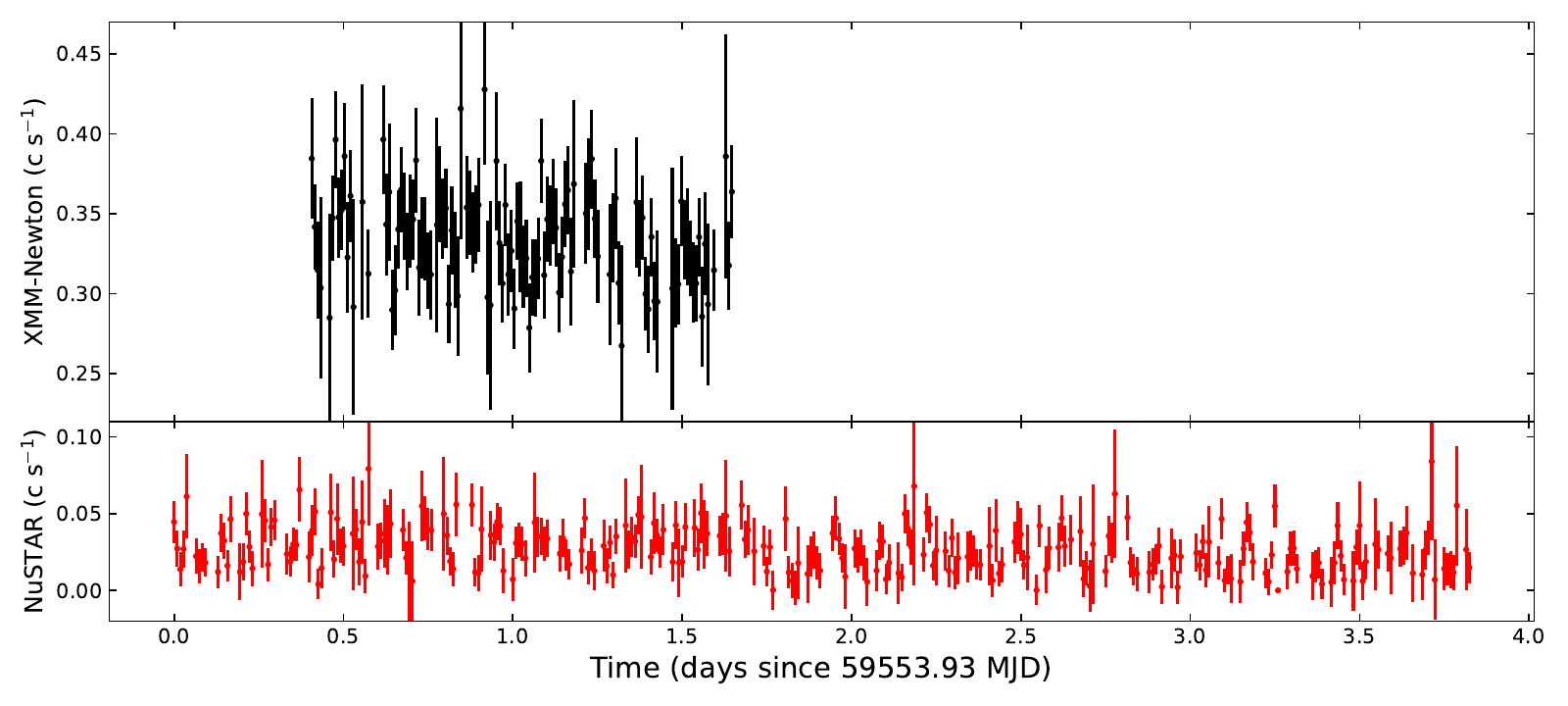}
    \captionsetup{font=small}
    \caption{Background-subtracted light curves of NGC 5055 X-1 extracted from the \xmm\ and \nustar~calibrated data in the 0.3–12 keV and 3–78 keV energy bands, respectively, both using a binning length of 800~s.}
    \label{fig:lcurve}
\end{figure}

We analyzed the background-subtracted light curves of \ngc using combined data from the \xmm\ and \nustar~observations to search for any significant variability. The light curves were extracted in the 0.3--12 keV energy range for \xmm\ and in the 3--78 keV range for \nustar. For each instrument, we visually inspected the light curves to search for high or low energy variability across several timescales.
For \xmm\ (pn camera), the light curve appears approximately constant, with an average count rate of ${\sim}0.33$~$\mathrm{counts~s^{-1}}$ (see \hyperref[fig:lcurve]{Fig~\ref{fig:lcurve}}). Similarly, the \nustar light curve shows no significant variability, with an average count rate of ${\sim}0.045$~$\mathrm{counts~s^{-1}}$. 

We used the spectral timing algorithms provided by the \texttt{HENDRICS} \citep{HENDRICS} and \texttt{Stingray} \citep{stingray2019ApJ...881...39H} software packages to investigate potential pulsations signals within the \xmm and \nustar~calibrated event files, using the same energy ranges as above. We performed this search in frequency ranges appropriate for each instrument, based on their sampling times and observation mode. 

For the \xmm\ EPIC-PN data, we searched for pulsations in the 0.01--6.81 Hz frequency range since the full-frame mode provides a Nyquist frequency of about 6.81 Hz. For \nustar, despite the shorter readout times of the FPMA/B detectors, which allow a theoretical Nyquist frequency of up to 200 Hz, we restricted our search to the 0.01--10.0 Hz range to avoid high-frequency noise and to be consistent with the \xmm\ analysis.

We used the \texttt{HENaccelsearch} tool to search for pulsation candidates in the 0.01--6.81 Hz range, allowing for variable period derivatives up to $10^{-9}$~Hz~s$^{-1}$. No significant pulsation candidates were detected in any of the datasets. We then applied the \texttt{HENz2vspf} tool to determine an upper bound on the pulse fraction (PF). The PF is defined as the ratio between the amplitude of the pulsation and the maximum flux, potentially exhibited by each signal. This procedure scrambles the event times and adds a pulsation with a random pulsed fraction. It then extracts the maximum of the $Z^2$ distribution within a small interval around the pulsation. This process is repeated $N_{\rm trial}$ times.
Based on 10000 simulations, we derived an upper limit for the PF (with 90\% confidence). Specifically, we obtain a PF upper limit of 10\%for the \xmm\ data and 32\% for the \nustar~data.

\subsection{Spectral analysis} \label{spec}

\renewcommand{\arraystretch}{1.45} 
\begin{table}[]
    \begin{center}
    \caption{Best-fit parameters with 1$\sigma$ level uncertainties derived from modeling the simultaneous \xmm~+\nustar dataset of \ngc.}
    \label{tab:table_spectra}
    \begin{adjustbox}{max width=\columnwidth}
    \begin{tabular}{l c c c c}
    \hline
    \hline
    Component       & Parameter                               & {\sc diskbb+diskpbb}            & {\sc cutoffpl}                  & {\sc simpl}   \\
    \hline
    {\sc cons}      & $\mathcal{C}_{\rm MOS1}$                & 1.03$\pm0.02$             & 1.03$\pm0.02$             & 1.03$\pm0.02$ \\ 
                    & $\mathcal{C}_{\rm MOS2}$                & 1.01$\pm0.02$             & 1.01$\pm0.02$             & 1.01$\pm0.02$ \\ 
                    & $\mathcal{C}_{\rm FPMA}$                & 0.89$\pm0.04$             & 0.89$\pm0.04$             & 0.89$\pm0.04$ \\
                    & $\mathcal{C}_{\rm FPMB}$                & 0.93$\pm0.04$             & 0.94$\pm0.04$             & 0.94$\pm0.04$ \\
    {\sc tbnew$\_$pcf} & N$_{\rm H}$ [10$^{22}$ cm$^{-2}$]    & 0.56$\pm0.05$             & 0.57$\pm0.05$             & 0.53$\pm0.05$ \\ 
                    & {\sc f$_{\rm cov}$}                     & 0.83$^{+0.03}_{-0.04}$    & 0.80$^{+0.04}_{-0.05}$    & 0.75$^{+0.04}_{-0.06}$  \\
    {\sc diskbb}    & {\sc T$_{\rm in}$} [keV]                & 0.18$\pm0.01$             & 0.19$\pm0.01$             & 0.20$\pm0.01$  \\
                    & {\sc norm}                              & 90$^{+50}_{-30}$          & 60$^{+40}_{-20}$          & 40$^{+20}_{-10}$  \\
    {\sc simpl}     & {\sc $\Gamma_{\rm s}$}                  & -                         & -                         & 3.0$^{+0.5}_{-0.6}$ \\ 
                    & {\sc CF}                                & -                         & -                         & 0.6$\pm0.3$ \\
    {\sc diskpbb}   & {\sc T$_{\rm in}$} [keV]                & 2.9$\pm0.2$               & 1.9$^{+0.3}_{-0.2}$       & 1.7$\pm0.2$ \\
                    & {\sc p}                                 & 0.56$\pm0.01$             & 0.59$^{+0.04}_{-0.03}$    & 0.62$\pm0.03$ \\
                    & {\sc norm} [$\times$10$^{-4}$]          & 1.8$^{+0.6}_{-0.5}$       & 9$^{+6}_{-4}$             & 17$^{+9}_{-6}$ \\          
    {\sc cutoffpl}  & {\sc $\Gamma_{\rm c}$}                  & -                         & 0.59$^{\dagger}$          & - \\    
                    & {\sc E$_{\rm cut}$} [keV]               & -                         & 7.1$^{\dagger}$           & - \\
                    & {\sc norm} [$\times$10$^{-6}$]          & -                         & 20$\pm4$                  & - \\ 
    \\
    {\sc cflux}     & {\sc F$_{\rm X}$} [10$^{-12}$ erg s$^{-1}$ cm$^{-2}$] & 2.02$^{+0.52}_{-0.41}$  & 1.9$^{+0.5}_{-0.4}$      & 1.8$^{+0.5}_{-0.3}$ \\                     
    {\sc cglumin}   & {\sc L$_{\rm X}$} [10$^{40}$ erg s$^{-1}$]            & 1.97$^{+0.51}_{-0.40}$  & 1.86$^{+0.48}_{-0.38}$   & 1.76$^{+0.51}_{-0.81}$ \\
    \\
    \hline
    \multicolumn{2}{c}{$\chi^2$/dof}                          & 447.2/467                       & 442.6/466                      & 444.3/465      \\
    \hline
    \end{tabular}
    \end{adjustbox}
    \end{center}
    \footnotesize{\textbf{Notes.} Reported fluxes and luminosities, which assume a distance of 9.04~Mpc, are computed in the 0.3--20~keV energy range. Parameters marked with ${}^{\dagger}$ were frozen during the fit.}
\end{table}

The X-ray spectral analysis of \ngc was performed using XSPEC v.12.13.1 \citep{1996ASPC..101...17A}. The spectra from \xmm (0.3--10 keV) and \nustar~(3--20 keV) were grouped to ensure a minimum of 20 counts per bin, allowing for the use of $\chi^2$ statistics. The energy ranges were chosen to avoid background-dominated regions above $10$ keV for \xmm and $20$ keV for \nustar. Unabsorbed fluxes and luminosities were derived using convolution models {\tt cflux} and {\tt cglumin}, assuming a distance of 9.04 Mpc \citep{Tully_2013,McQuinn_2017, Karachentsev_2020}. To account for cross-calibration uncertainties, multiplicative constants were included in all spectral fits. The constant for EPIC-pn was fixed at 1, while those for MOS1, MOS2, and \nustar~FPMA/FPMB were allowed to vary. The difference between MOS1 and MOS2 was $\sim5\%$, and the difference between FPMA/FPMB was $\sim10\%$, both relative to EPIC-pn \citep[see e.g.,][]{Saavedra2023A&A...680A..88S}. The broadband spectrum was modeled simultaneously in the 0.3--20 keV range and residuals were carefully examined to assess the quality of the fit. 

Uncertainties were estimated using Markov chain Monte Carlo (MCMC) chains, generating $10^6$ samples using the {\tt chain} command in XSPEC. The number of walkers was set to $12$ times the number of free parameters for each model \citep[see][for more details]{Cruz_2025}. The convergence of the MCMC chains was verified by inspecting the trace plots of each parameter and ensuring that the chains had reached a stable state. This was confirmed by checking that the autocorrelation time was small (close to unity), indicating that the walkers had efficiently explored the parameter space (see \citealt{Fogantini2023, saavedragx13} for more details). 

To classify the spectral state of \ngc, we followed the scheme proposed by \citet{Sutton2013MNRAS}, applying a model of the form \texttt{tbnew}\footnote{https://pulsar.sternwarte.uni-erlangen.de/wilms/research/tbabs/}*\texttt{tbnew}*(\texttt{diskbb}+\texttt{powerlaw}) in the 0.3--10~keV energy range, adopting the photoionization cross sections of \citet{1996ApJ...465..487V} and the elemental abundances of \citet{2000ApJ...542..914W}. In this model, the \texttt{diskbb} component \citep{mitsuda1984PASJ...36..741M} accounts for thermal emission from a multicolor accretion disk, while the \texttt{powerlaw} component represents a hard, nonthermal tail likely produced by Compton scattering or optically thin coronal emission. This model forms the basis of the decision-tree classification diagram introduced by \citet{Sutton2013MNRAS}, which relies exclusively on \xmm data.

\begin{figure}
\centering
    \includegraphics[width=\columnwidth]{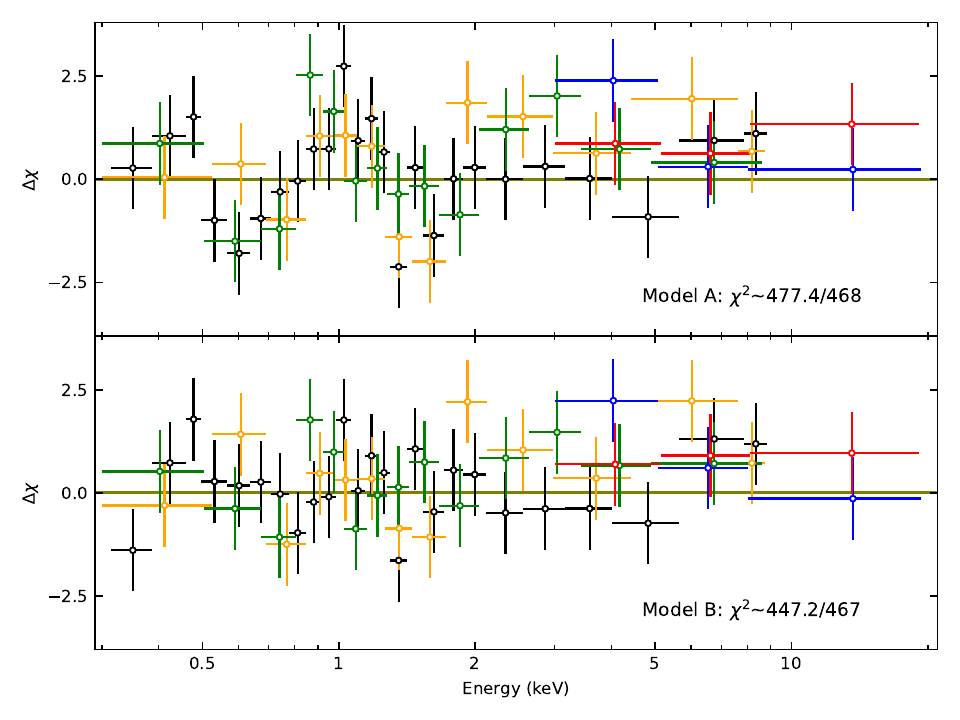}
    \captionsetup{font=small}
    \caption{Residuals from fitting two absorption configurations to the joint \xmm and \nustar~spectra (0.3--20~keV). \textit{Top panel:} Model A = \texttt{tbnew*(diskbb+diskpbb)} (single neutral absorber; Galactic column only), which reveals a feature around ${\sim}0.95$~keV. \textit{Bottom panel:} Model B = \texttt{tbnew*tbnew\_pcf*(diskbb+diskpbb)} (additional partial-covering neutral absorber), in which this feature is unnoticeable. Data have been rebinned for visualization only.}
    \label{fig:res}
\end{figure}

\begin{figure}
\centering
    \includegraphics[width=\columnwidth]{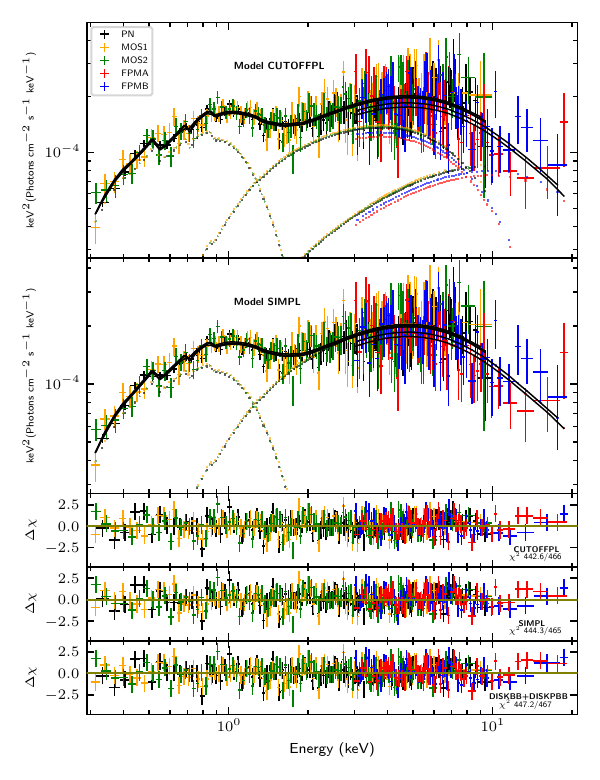}
    \captionsetup{font=small}
    \caption{\xmm+ \nustar~unfolded spectra of \ngc. The top panel shows the best-fit {\sc cutoffpl} model. The second panel displays the {\sc simpl} model fit. The bottom three panels show the residuals for each tested scenario with their corresponding $\chi^2$ values: {\sc cutoffpl} (third panel), {\sc simpl} (fourth panel), and {\sc diskbb+diskpbb} (fifth panel).}
    \label{fig:fit}
\end{figure}

To remain consistent with the methodology of \citet{Sutton2013MNRAS}, we repeated the former procedure using only \xmm data. In this case, the model provided a substantially improved fit, with a reduced $\chi^2$ of 1.07 for 345 d.o.f. The best-fit parameters were a disk temperature of kT$_{\rm in}$=0.24$\pm0.01$~keV and photon index of $\Gamma$=1.66$\pm0.06$, both consistent with sources classified in the HUL regime. Based on these results and the decision-tree criteria of \citet{Sutton2013MNRAS}, we classify \ngc as being in the HUL state.

We modeled the 0.3--20~keV broadband X-ray spectrum of \ngc\ with a two-layer neutral absorber along the line of sight, using the composite model \texttt{tbnew*tbnew\_pcf}. An initial fit with the standard \texttt{tbabs} left a prominent residual around 0.95~keV. We therefore employed \texttt{tbnew} for its improved treatment of low-energy edges; however, the 0.95~keV feature persisted. Introducing an additional partial-covering neutral absorber, \texttt{tbnew\_pcf}, removed this structure (see Fig.~\ref{fig:res}) and significantly improved the fit. In our final configuration, the \texttt{tbnew} component accounts for the Galactic column, fixed at $N_{\mathrm{H}}=3.72\times10^{20}\ \mathrm{cm^{-2}}$ following the H\,\textsc{i} survey of \citet{Kalberla2005A&A}, while \texttt{tbnew\_pcf} represents intrinsic neutral absorption with partial covering -- plausibly due to clumpy material or a fragmented disk wind near the compact object -- with its column density and covering fraction left free.

To explore the nature of the compact object in \ngc, we tested two physical scenarios commonly invoked in the ULX population. The first assumes a nonmagnetic stellar-mass BH accreting above the Eddington limit, while the second considers a strongly magnetized NS, as in the case of PULXs, despite the absence of detected pulsations in this source (see Section~\ref{lcur}). In both cases, the joint \xmm and \nustar~spectra were used to model the continuum, incorporating the same two-component neutral absorption structure described earlier.

In the nonmagnetic scenario, we adopted a baseline, thermal model consisting of two multicolor accretion-disk components: a cooler outer disk represented by \texttt{diskbb} and a hotter inner disk described by \texttt{diskpbb}. To account for the Comptonization of the inner-disk photons, we convolved the hotter component with the \texttt{simpl} model \citep{Steiner2009PASP..121.1279S}. The complete model was \texttt{tbnew*tbnew\_pcf*(diskbb+simpl*diskpbb)}, hereafter referred to as the \texttt{simpl} model. The \texttt{diskpbb} component \citep{Mineshige1994ApJ...435L.125M} includes a free radial temperature index, $p$, where T(r) $\propto$ r$^{-p}$. While $p$ = 0.75 corresponds to a standard thin disk, lower values ($p\lesssim0.7$) are indicative of advection-dominated (slim) flows typical of super-Eddington accretion.

In the magnetized accretor scenario, we adopted the same thermal disk structure but replaced the Comptonization component with a high-energy tail modeled by a cutoff power law (\texttt{cutoffpl}). This component is commonly used to describe emission from magnetically collimated accretion columns \citep{Walton2018MNRAS.473.4360W, Walton_2018}. The photon index and cutoff energy were fixed at $\Gamma = 0.59$ and E$_{\rm cut}$=7.1~keV \citep[following:][]{Brightman2016ApJ...816...60B, Walton_2018}. The resulting configuration, \texttt{tbnew*tbnew\_pcf*(diskbb+diskpbb+cutoffpl)}, is hereafter referred to as the \texttt{cutoffpl} model.

Both models yielded statistically acceptable fits to the spectrum over the 0.3--20~keV energy range, with similar $\chi^2/{\rm dof}$ and comparable residuals (see \hyperref[tab:table_spectra]{Table~\ref{tab:table_spectra}} and \hyperref[fig:fit]{Fig.~\ref{fig:fit}}). 
The intrinsic absorption was $N_{\mathrm{H}}\sim5\times10^{21}$~cm$^{-2}$ with covering fractions of $f_{\rm cov}\sim0.8$. 
In all cases the continuum is dominated by a cool outer disk at $kT\sim0.19$~keV plus a hotter inner flow at $kT\sim1.8$--$2.1$~keV, with $p\sim0.6$. 
The {\sc simpl} model requires a steep Compton tail with $\Gamma_{\rm s}\sim3.3$ and a scattering fraction of $\sim70\%$, while the {\sc cutoffpl} component contributes $\lesssim11\%$ of the total flux. 
Quantifying the preference for the additional high-energy terms using the likelihood-ratio test in \textsc{xspec} with $10^{4}$ simulations, we found that adding {\sc simpl} is favored at $2.4\sigma$ and adding {\sc cutoffpl} at $2.2\sigma$. 
Therefore, a purely thermal {\sc diskbb+diskpbb} model provides an adequate description of the data without requiring a statistically significant high-energy excess (see \hyperref[tab:table_spectra]{Table~\ref{tab:table_spectra}}).
In all cases, the unabsorbed luminosity in the 0.3--20~keV band is $\sim2\times10^{40}$~erg~s$^{-1}$.
For completeness, Appendix~\ref{app:mcmc} presents the two-dimensional posterior projections of the MCMC results for the {\tt cutoffpl} and {\tt simpl} models (Figs.~\ref{fig:mcmc_cutoffpl} and \ref{fig:mcmc_simpl}).
These results establish a robust spectral characterization of NGC~5055~ULX~X--1 under both accretion scenarios, setting the foundation for further interpretation.

\section{Discussion} \label{sec:discussion}

\subsection{Spectral state and inner-flow geometry}
\label{sec:geom_state}

\ngc\ radiates at an unabsorbed 0.3--20 keV luminosity of ${\sim}2{\times}10^{40}\,\mathrm{erg\,s^{-1}}$, firmly in the ULX regime where super-Eddington accretion and/or mild beaming are required to explain the observed output. The measured fluxes, $F_{0.3\text{--}1.5~{\rm keV}}=1.0\times10^{-12}\,\mathrm{erg\,cm^{-2}\,s^{-1}}$ and $F_{1.5\text{--}10~{\rm keV}}=5.5\times10^{-13}\, \mathrm{erg\,cm^{-2}\,s^{-1}}$, give a hardness ratio of $H_{0.3\text{--}10~{\rm keV}}=0.53$, while the color--color ratios ${\rm S}=F_{2\text{--}4~{\rm keV}}/F_{4\text{--}6~{\rm keV}}=1.57$ and $H=F_{6\text{--}30~{\rm keV}}/F_{4\text{--}6~{\rm keV}}=1.7$ locate the source on the left side of the hardness--luminosity diagram (HLD; \citealp{Gurpide2021AaAa}) -- between the $10\times$ and $100\times$ Eddington tracks for a $1.4\,M_{\odot}$ NS -- and in the central region of the color--color plane (CCD; \citealp{Pintore_2017}) that hosts the majority of non-PULXs. These diagnostics place \ngc\ well away from the hard, highly variable branch that typically contains PULXs, suggesting a geometry dominated by an inflated inner disk and a dense, optically thick wind rather than by an energetically dominant accretion column. 

Over the 0.3–20 keV band, the broadband continuum is naturally captured by a nonmagnetic configuration in which a cool multicolour disk (\texttt{diskbb}) feeds a hotter, radially stratified inner flow (\texttt{diskpbb}). Mild Compton up-scattering in a warm, optically thick corona (\textsc{simpl}) can imprint the smooth curvature below $\sim$10 keV and any faint, steep nonthermal tail, while an additional cutoff power law (\textsc{cutoffpl}) may represent a weak hard excess if present. Crucially, these extra high-energy terms are not required by the data: a two-temperature thermal spectrum (\texttt{diskbb+diskpbb}), i.e., a combination of blackbody-like components, already provides an adequate description. This spectral morphology -- a hard, disk-dominated continuum with a broadened inner flow and only a weak high-energy tail -- is characteristic of the HUL regime, as seen in archetypal HUL ULXs such as NGC~1313~X-1 and Holmberg~IX~X-1 \citep{Sutton2013MNRAS, LuangtipMNRAS1282, Pinto2020MNRAS.492.4646P}, and contrasts with confirmed ULX pulsars, whose pulsed high-energy continua are typically described by hard cutoff power laws rather than by a steep power-law excess \citep[e.g.,][]{Brightman2016ApJ...829...28B,Walton_2018}.

The presence of partial covering absorption with a high covering fraction (f$_{\rm cov} \sim 0.8$) suggests a structured and optically thick absorber close to the accreting source. Such conditions are naturally expected in super-Eddington accretion regimes, where geometrically thick inflows and radiatively driven winds can obscure the inner emission \citep[see e.g.,][]{2023A&A...671A...9A, Combi2024A&A...686A.121C}. Similar absorption features have been reported in ULXs such as NGC~1313~X-1 and Holmberg~IX~X-1, and are often interpreted as clumpy winds or warped structures in the inner accretion flow \citep[][]{2016Natur.533...64P,Pinto2020MNRAS.492.4646P,Kosec2018MNRAS.473.5680K, Kosec2021MNRAS.508.3569K}. At the CCD-resolution of \xmm, such clumpy winds frequently manifest as an apparent excess/feature near $\sim$1~keV, very similar to that seen here, consistent with the low-resolution spectra compiled by \citet{Middleton2015MNRAS.454.3134M} for sources in which high-resolution studies reveal powerful winds. In our case, the inclusion of the \texttt{tbnew\_pcf} component not only improves the spectral fit significantly, but also provides strong evidence of a complex and anisotropic environment around the compact object.

The timing analysis reveals no coherent pulsations, with a $10\%$ pulsed-fraction upper limit. However, such non-detection does not allow us to discard the idea that the compact object has a neutron-star nature: some ULX pulsars show very low pulsed fractions (e.g., NGC~1313~X-2 at $\sim$5\%; \citealt{Sathyaprakash_2019}) and pulsations can be transient or intermittent (e.g., M82~X-2; \citealt{Bachetti2020ApJ...891...44B}). The source's position in the CCD and HLD diagrams, as well as the disk-like continuum, are all consistent with a black-hole accretor. Although a neutron-star accretor cannot be ruled out, we explore the black-hole scenario in detail to obtain an estimation of its putative mass.

\subsection{Black-hole mass estimation}
\label{sec:BH_mass}

Adopting the best–fit configuration for a stellar-mass black-hole accretor, we describe the time-averaged 0.3–20 keV spectrum with the composite model \texttt{tbabs*(diskbb + simpl*diskpbb)}.  In this framework, the \texttt{diskbb} component represents the quasi-standard outer disk, while \texttt{diskpbb} captures the advection-dominated inner flow through a variable radial temperature index, $p$.  The \texttt{simpl} convolution then models Compton up-scattering of seed photons in a tenuous, hot corona, producing the
observed high-energy tail.

The physical scale of the accretion flow is derived from the \texttt{diskpbb} normalization \citep{Kubota1998PASJ...50..667K, Makishima1988Natur.333..746M}:
\begin{equation}
N_{\mathrm{pbb}} = \left(\frac{r_{\mathrm{app}}/\mathrm{km}}{D_{10}}\right)^{2}\cos\theta,
\label{eq:Npbb}
\end{equation}
where $r_{\mathrm{app}}$ represents the apparent inner radius, $D_{10}=D/(10~\mathrm{kpc})$ is the distance scaling, and $\theta$ denotes the inclination angle. For super-Eddington accretion regimes, the disk becomes geometrically thick, with the locally emitted spectrum hardened by electron scattering \citep{Watarai2003PASJ...55..959W, Kawaguchi2003ApJ...593...69K, Shrader2003ApJ...598..168S, Isobe2012PASJ...64..119I}. Following \citet{Vierdayanti2008PASJ...60..653V}, we applied corrections to the apparent inner radius:
\begin{equation}
R_{\mathrm{in}} = \xi\kappa^{2}r_{\mathrm{app}},
\label{eq:Rin}
\end{equation}
adopting a geometric correction factor, $\xi = 0.353$, and a color-hardening factor, $\kappa = 3.0$. The black-hole mass implied by the uncorrected slim disk formulation is
\begin{equation}
M_{X} = \frac{R_{\mathrm{in}}}{R_{\mathrm{ISCO}}}, \quad 
R_{\mathrm{ISCO}} = 6\,\dfrac{G M_{\odot}}{c^{2}} = 8.9~\mathrm{km},
\label{eq:Mx}
\end{equation}
where $R_{\mathrm{ISCO}}$ is the innermost stable circular orbit for a nonrotating BH. Since advection shifts the disk's inner edge approximately 20\% inside $R_{\mathrm{ISCO}}$, we calculated the black-hole mass to be
\begin{equation}
M_{\mathrm{BH}} = 1.2M_{X}.
\label{eq:MBH}
\end{equation}

\noindent Using a distance to the source of $9.04~\mathrm{Mpc}$, a face-on accretion disk ($\theta = 0$), and the best-fit normalization of the {\sc diskpbb} component, {\bf $N_{\mathrm{pbb}} = (17^{+9}_{-6})\times10^{-4}$}, Equations~(\ref{eq:Npbb})--(\ref{eq:MBH}) yield a nominal black-hole mass of {\bf $M_{\mathrm{BH}} \approx 16.0~M_{\odot}$. }
 To assess the impact of systematic effects, we carried out a Monte-Carlo simulation with \(2\times10^{5}\) realizations using priors: the distance was modeled as a truncated Gaussian with a mean of \(9.04~\mathrm{Mpc}\) and standard deviation of \(0.40~\mathrm{Mpc}\); the inclination was drawn from a uniform distribution in \(\cos\theta\in[0.3,1.0]\), as expected if high-inclination sight lines are preferentially obscured by thick winds/funnel structures \citep[e.g.,][]{Sutton2013MNRAS,Middleton2015MNRAS.454.3134M,2016Natur.533...64P}; the geometric factor adopted a truncated Gaussian centered at \(\xi=0.353\) with \(\sigma=0.050\), appropriate for slim disks \citep[cf. thin-disk \(\xi\simeq0.41\) in][]{Vierdayanti2008PASJ...60..653V,Kubota1998PASJ...50..667K}; and the color–hardening factor was sampled uniformly from \(\kappa\in[2.0,3.5]\), spanning values expected at high accretion rates \citep[e.g.,][]{Kawaguchi2003ApJ...593...69K,Isobe2012PASJ...64..119I,Davis2005ApJ...621..372D,Shimura1995ApJ...445..780S}. Under these assumptions we obtained {\bf $M_{\mathrm{BH}}=17.5^{+8.5}_{-6.2}\,M_{\odot}$ ($1\sigma$).}
Our mass determinations are consistent with the hypothesis that the compact object is a stellar-mass BH accretor.
\section{Conclusions}\label{sec:conclusions}

We have carried out the first analysis of a simultaneous \xmm\ and \nustar\ observing campaign on the ULX \ngc, achieving $\sim$100~ks of effective exposure and continuous spectral coverage from 0.3 to 20~keV.  No coherent pulsations are detected in the light curves of pn and FPM instruments, with $90\%$ upper limits on the pulsed fraction of 10\,\% (\xmm) and 32\,\% (\nustar).  The time-averaged spectrum is dominated by a cool outer disk together with a hotter, advective inner flow, partially obscured by neutral material. The unabsorbed broadband luminosity, $L_{0.3-20\,\mathrm{keV}}\simeq2\times10^{40}\,\mathrm{erg\,s^{-1}}$, places \ngc\ firmly within the ultraluminous regime.  Hardness-ratio and color–color diagnostics locate the source on the hard-ultraluminous branch, a region typically associated with super-Eddington accretion funneled through an inflated inner disk and an optically thick wind.  After correcting the radial-temperature-profile normalization for color and geometric factors, we infer a black-hole mass of {\bf $M_{\mathrm{BH}}=17.5^{+8.5}_{-6.2}\,M_{\odot}$. } This mass estimate was derived under the assumption of a black-hole accretor; in the absence of definitive neutron-star signatures, a neutron-star accretor cannot be ruled out. High-resolution X-ray spectroscopy and long-baseline timing observations will be essential to detect wind signatures, constrain the viewing geometry, and search for any intermittent pulsations at lower flux levels.

\begin{acknowledgements}
We thank the anonymous reviewer for their valuable comments on this manuscript. Funded by the European Union (Project 101183150 - OCEANS). Views and opinions expressed are however those of the author(s) only and do not necessarily reflect those of the European Union or the European Research Executive Agency (REA). Neither the European Union nor REA can be held responsible for them.  EAS acknowledges support by the Spanish \textit{Agencia estatal de investigaci\'on} via PID2021-124879NB-I00. FAF is a postdoc fellow of CONICET. JAC and FG are CONICET researchers. FAF, JAC, and FG acknowledge support by PIP 0113 (CONICET). FG acknowledges support from PIBAA 1275 (CONICET). JAC was supported by Consejería de Economía, Innovación, Ciencia y Empleo of Junta de Andalucía as research group FQM-322. FG and JAC were also supported by grant PID2022-136828NB-C42 funded by the Spanish MCIN/AEI/ 10.13039/501100011033 and “ERDF A way of making Europe”. 
\end{acknowledgements}

\bibliographystyle{aa}
\bibliography{biblio}

\appendix
\twocolumn
\section{Posterior distributions from the spectral models}\label{app:mcmc}

\begin{figure*}[!ht]
\centering

\begin{subfigure}{0.5\textwidth}
\includegraphics[width=\textwidth]{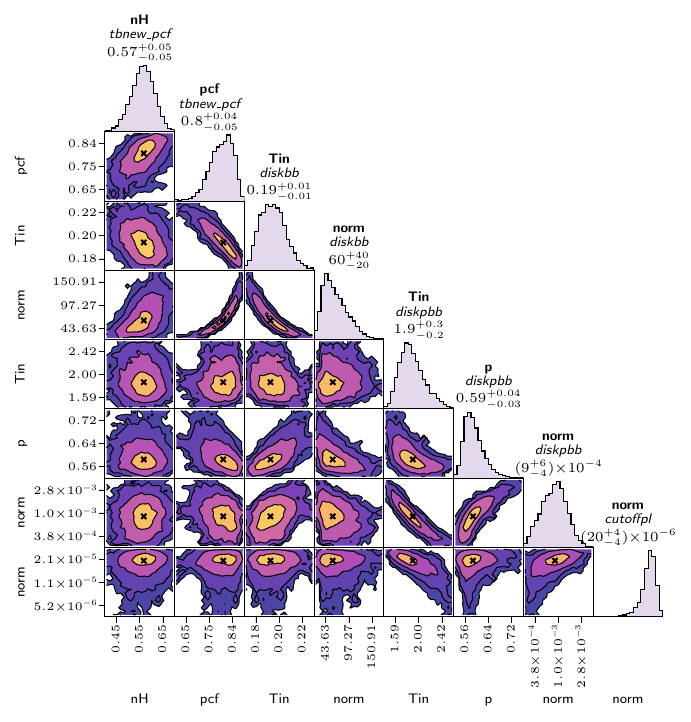}
\caption{{\tt cutoffpl}}
\label{fig:mcmc_cutoffpl}
\end{subfigure}\hfill
\begin{subfigure}{0.5\textwidth}
\includegraphics[width=\textwidth]{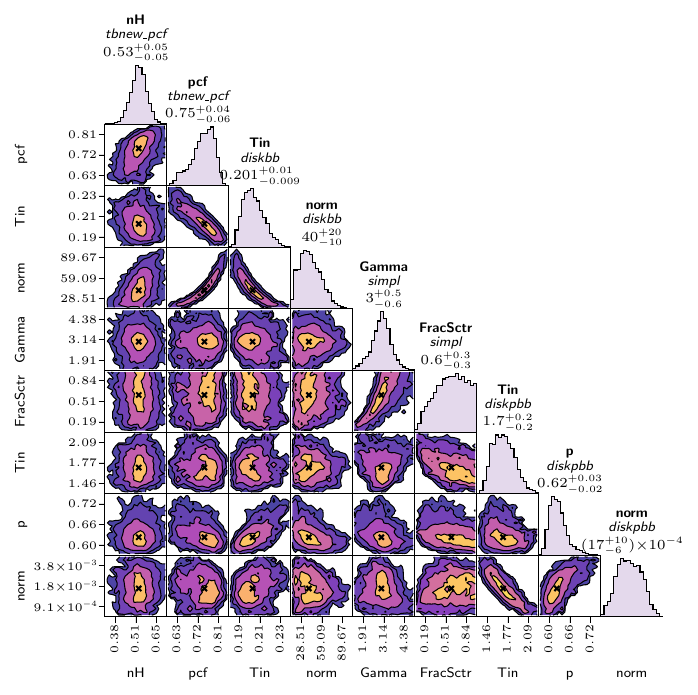}
\caption{{\tt simpl}}
\label{fig:mcmc_simpl}
\end{subfigure}

\vspace{2mm}

\begin{subfigure}{0.5\textwidth}
\centering
\includegraphics[width=\textwidth]{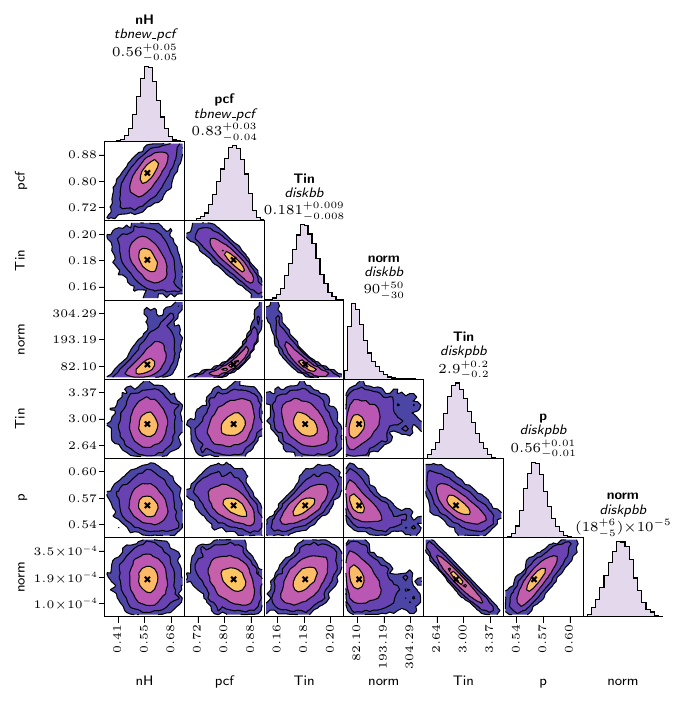}
\caption{{\tt diskbb+diskpbb}}
\label{fig:mcmc_diskbb}
\end{subfigure}

\caption{\small
Corner plots showing the posterior probability distributions of the best-fit parameters obtained from the MCMC analysis of the three tested models: {\tt cutoffpl} (top left), {\tt simpl} (top right), and {\tt diskbb+diskpbb} (bottom). 
N$_{\rm H}$ and T$_{\rm in}$ are expressed in units of $10^{22}$~cm$^{-2}$ and keV, respectively. 
Each distribution is based on $10^{6}$ converged MCMC iterations.}
\label{fig:mcmc_all}
\end{figure*}

We present here the posterior probability distributions obtained from the MCMC analysis of the three spectral models tested in this work: {\tt cutoffpl}, {\tt simpl}, and {\tt diskbb+diskpbb}. Figures~\ref{fig:mcmc_cutoffpl}, \ref{fig:mcmc_simpl}, and \ref{fig:mcmc_diskbb} show the corresponding corner plots, which display the one- and two-dimensional marginalized distributions of the main free parameters. 
Each plot is based on $10^{6}$ converged MCMC iterations.

\twocolumn 

\end{document}